\documentclass{article}
\usepackage[utf8]{inputenc}
\usepackage{amsmath}
\makeatletter
\def\tagform@#1{\maketag@@@{[\ignorespaces#1\unskip\@@italiccorr]}}
\makeatother
\usepackage{bm}
\usepackage{authblk}
\usepackage{amssymb}
\usepackage{graphicx}
\DeclareGraphicsExtensions{.png,.pdf,.eps}
\usepackage{ifpdf}
\usepackage{cite}
\usepackage{doi}

\usepackage{tikz}
\usepackage[
	capitalise,
]{cleveref}

\usepackage{etoolbox}
\usepackage{color}

\newtoggle{MRM}

\togglefalse{MRM}

\iftoggle{MRM}{
\ifpdf{
	\typeout{using pdf(la)tex}
	\usepackage[]{hyperref}
\else
	\typeout{using plain (la)tex}
	\usepackage[dvipdfm]{hyperref}
\fi
}
\usepackage[nomarkers, nolists]{endfloat}
}

\usepackage{changes}
\newcommand{\stkout}[1]{\ifmmode\text{\sout{\ensuremath{#1}}}\else\sout{#1}\fi}
\setdeletedmarkup{\stkout{#1}}

\usepackage{ifdraft}
\usepackage{marginnote}

\usepackage[
per-mode=fraction,
separate-uncertainty,
retain-unity-mantissa=false,
product-units=power,
table-alignment=center,
detect-all,
binary-units=true,
exponent-product=\cdot,
exponent-base=10
]{siunitx}

\newlength\Colsep
\newlength{\savedparindent}

\title{Frequency-modulated SSFP with radial sampling and subspace 
reconstruction: A time-efficient alternative to phase-cycled bSSFP}
\author[1]{Volkert Roeloffs}
\author[2,3]{Sebastian Rosenzweig}
\author[2,3]{H. Christian M. Holme}
\author[2,3]{Martin Uecker}
\author[1,3]{Jens Frahm}

\affil[1]{Biomedizinische NMR Forschungs GmbH am Max-Planck-Institut für biophysikalische Chemie, 37070 Göttingen, Germany}
\affil[2]{Institute for Diagnostic and Interventional Radiology,	University Medical Center, 37075 Göttingen, Germany}
\affil[3]{German Centre for Cardiovascular Research (DZHK), partner site Göttingen, Germany}

\begin{document}
\setlength{\savedparindent}{\parindent}

\maketitle

\vfill

\noindent
{\em Running head:} {Frequency-modulated SSFP and subspace reconstruction}

\vspace{0.5cm}

\noindent
{\em Correspondence to:} \\
Dr. V. Roeloffs\\
Biomedizinische NMR Forschungs GmbH\\
am Max-Planck-Institut für biophysikalische Chemie\\
37070 Göttingen, Germany\\
\texttt{volkert.roeloffs@mpibpc.mpg.de}

\vspace{0.3cm}
\noindent
Approximate word count: 223 (abstract) 5300 (body)\\
\iftoggle{MRM}{
Number of pages: 31\\
Number of figures: 8\\
Number of tables: 0\\
}

\vspace{0.3cm}
\noindent
Submitted to \textit{Magnetic Resonance in Medicine} as a Full Paper\\
\begin{tabular}{@{}lc}
Date of submission: & {2018-03-16} \\
Date of revision: &{2018-07-05}
\end{tabular}

\vspace{0.3cm}
\noindent

\newpage
\setlength{\parindent}{0in}
\section*{Abstract}

{\bf Purpose:} 
A novel subspace-based reconstruction method for frequency-modulated balanced steady-state free precession (fmSSFP) MRI is presented. In this work, suitable data acquisition schemes, subspace sizes, and efficiencies for banding removal are investigated.

{\bf Theory and Methods:}
By combining a fmSSFP MRI sequence with a 3D stack-of-stars trajectory, scan efficiency is maximized as spectral information is obtained without intermediate preparation phases. A memory-efficient reconstruction routine is implemented by introducing the low-frequency Fourier transform as a subspace which allows for the formulation of a convex reconstruction problem. The removal of banding artifacts is investigated by comparing the proposed acquisition and reconstruction technique to phase-cycled bSSFP MRI. Aliasing properties of different undersampling schemes are analyzed and water/fat separation is demonstrated by reweighting the reconstructed subspace coefficients to generate virtual spectral responses in a post-processing step.

{\bf Results:} 
A simple root-of-sum-of-squares combination of the reconstructed subspace coefficients yields high-SNR images with the characteristic bSSFP contrast but without banding artifacts. Compared to Golden-Angle trajectories, turn-based sampling schemes were superior in minimizing aliasing across reconstructed subspace coefficients. Water/fat separated images of the human knee were obtained by reweighting subspace coefficients.

{\bf Conclusion:}
The novel subspace-based fmSSFP MRI technique emerges as a time-efficient 
alternative to phase-cycled bSSFP. The method does not need intermediate 
preparation phases, offers high SNR and avoids banding artifacts. Reweighting 
of the reconstructed subspace coefficients allows for generating virtual 
spectral responses with applications to water/fat separation.

\vspace{0.5in}
{\bf Key words:} {subspace reconstruction; bSSFP; fmSSFP; frequency-modulated SSFP, banding, preparation phases}
\newpage

\setlength{\parindent}{\savedparindent}

\section*{Introduction}

The efficient use of scan time has been a permanent demand since the inception of MRI. Fast steady-state sequences based on principles of balanced steady-state free precession (bSSFP) offer the advantage of high SNR and excellent tissue-fluid contrast, but suffer from strong sensitivity to local off-resonances and magnetic field inhomogeneities which result in signal void in the form of banding artifacts.

To overcome these drawbacks, several methods have been proposed that employ specific schedules of transmit and receive phase modifications \cite{Schwenk_1971,Zur1990,Casselman1993,Bangerter_2004}. These techniques exploit the fact that the spectral response of a bSSFP sequence can be shifted by introducing linear phase increments from pulse to pulse. Images with removed or reduced bandings can then be reconstructed by combining multiple data sets acquired with different increments (phase cycling). However, each of these data sets demands the spin system to be in steady state. Consequently, each phase-cycled acquisition is accompanied by preceding transient phases without data acquisition that partly counterbalance the efficiency of bSSFP MRI.

With frequency-modulated SSFP (fmSSFP) \cite{Foxall2002} an acquisition technique was introduced that completely eliminates the need for intermediate preparation phases when imaging at multiple offset frequencies. Instead, the spectral offset frequency is slowly modulated in a linear fashion throughout the whole acquisition period resulting in a dynamic steady-state signal. In contrast to multiple-offset bSSFP, fmSSFP allows for continuous data acquisition and subsequent reconstruction of banding-free images by combining all acquired data and hence effectively averaging over the entire bSSFP frequency response profile. In combination with radial sampling and gridding reconstruction, fmSSFP has proven to provide banding-free images in reduced scan time when compared to bSSFP  \cite{Benkert2015a,Benkert_2016}. However, complex averaging of the spectral content results in a gradient spoiled contrast with accordingly reduced SNR when compared to on-resonant bSSFP.

Here, a memory-efficient subspace-based reconstruction of the full frequency response profile is proposed by combining fmSSFP imaging with a 3D stack-of-stars trajectory. Banding-free images with high SNR can be obtained by a simple root-of-sum-of-squares (RSS) combination of the reconstructed subspace coefficients. In contrast to multiple-offset bSSFP, where the frequency response profile is sampled only at a few offset frequencies, the fmSSFP acquisition provides full access to the frequency response profile, restricted only by the chosen subspace. This spectral information can be exploited to perform water/fat separation by designing a virtual spectral response similar to linear combination steady-state free precession (LCSSFP, \cite{Vasanawala_2000}).

The purpose of this work was to assess suitable subspace sizes and sampling schemes, to investigate  parameter dependencies of the generated image contrast, and to evaluate feasibility and robustness of the proposed fmSSFP reconstruction method in vitro and in vivo.

\section*{Theory}

\subsection*{Sequence}

Phase-cycled bSSFP or fmSSFP MRI sequences are generally characterized by balanced net gradient moments for each axis. Different spectral responses can be obtained by varying the phase of the radiofrequency (RF) excitation and adapting the receiver phase correspondingly. To characterize these RF phase modulation schemes for phase-cycled bSSFP or fmSSFP, it is sufficient to consider a second order polynomial of the form

\begin{align}
\phi(n)=\phi_0+\Delta\phi \;n+\Psi/2 \;n^2\ \text{,}
\tag{1}
\end{align}

where $\phi\left(n\right)$ denotes the transmit and receive phase for the $n$th pulse, $\phi_0$ a constant phase offset, $\Delta \phi$ a (linear) phase increment, and $\Psi$ a quadratic phase increment. While $\phi_0$ only rotates the frame of reference without affecting the measured signal, individual phase-cycled bSSFP images can be obtained by setting $\Psi=0^{\circ}$ and using different linear phase increments $\Delta \phi$  (or, equivalently, different constant offset-frequencies). The classic on-resonant, high-intensity signal for bSSFP is obtained with $\Delta\phi=180^{\circ}$.

In \cite{Foxall2002}, a certain tolerance of the bSSFP steady-state with respect to small and continuous changes in this linear increment was discovered. These changes can be described by small values of $\Psi$. Note that phase differences (modulo $360^{\circ}$) between individual pulses become a periodic function with period $360^{\circ} / \Psi$. Throughout this work a value of $\Psi=360^{\circ}/N$ was used, where $N$ is the total number of projections in a 3D MRI acquisition. In this manner, all phase differences are covered exactly once corresponding to a full sweep through the spectral response profile.

\subsection*{Subspace}

The idea behind choosing a subspace is to constrain the image reconstruction problem in such a way that it remains over-determined even when severely undersampled \cite{Huang2012,Petzschner_2011,Huang_2012,Pedersen2009,Zhao_2014,Tamir2016}. The size or dimensionality of a suitable subspace should be small, while still allowing for accurate signal representation. If these conditions are met, the degrees of freedom in the subspace-constraint reconstruction problem are significantly reduced compared to the signal representation in the original problem. This reduction allows to better recover overlapping signals in an aliased object and makes subspace-constraints appealing when reconstructing undersampled data sets that exhibit a high degree of aliasing but simple temporal dynamics \cite{Petzschner_2011}. To this end, a linear subspace constraint is particularly attractive for two reasons. First, linearity results in the formulation of a convex problem, and second, in the context of MRI, linearity inherently covers partial volume effects as linear combinations of signals in the subspace. A suitable linear basis for such a subspace can for example be found by principal component analysis of measured training signals \cite{Pedersen2009} or signals simulated for a known range of abundant tissue parameters \cite{Tamir2016,McGivney2014,Zhao_2014,Huang2012}.

Here, the low-frequency Fourier transform is chosen as a subspace. This choice is motivated by the rapid decay in magnitude of the bSSFP configuration modes with increasing order \cite{Hilbert_2017,Zur1990,Nguyen2016,Jung_2010}. As these modes (see \cite{zur1988,mizumoto1991,hanicke2003,Heule2014} for analytical expressions) reflect the Fourier coefficients of a bSSFP spectrum, the low-frequency subspace represents spectra up to a maximum frequency. However, for this work, the most import property of this particular subspace is given by the shift theorem of the Fourier transform, which guaranties that time shifts of the signal response (as caused by local off-resonances) translate to linear phase terms in the subspace. As a consequence, the accuracy of the subspace approximation remains independent of local off-resonance contributions and possible modelling errors are merely a function of T1, T2, and flip angle.

\subsection*{Reconstruction}

After selecting the subspace transform, the reconstruction problem can be formulated as an inverse problem with the forward model given by 

\begin{align}
y=\mathcal{P}_{\vec{k}}\mathcal{F}_{s}S\mathcal{F}_{t}^{-1}x
\tag{2}
\end{align}

where $y$ denotes the radial raw data from one full cycle of the fmSSFP measurement, $\mathcal{P}_{\vec{k}}$ the projection onto the sampled k-space trajectory, $\mathcal{F}_{s}$  the spatial Fourier transform, $S$ multiplication with the (predetermined) coil sensitivity profiles,  $\mathcal{F}_{t}$ the low-frequency temporal Fourier basis, and $x$ the unknown subspace coefficients.

Aligning the individual partitions allows for slice decoding in $k_z$ direction prior to image reconstruction by 1D inverse Fourier transform. This has the advantage of splitting the large 3D data set, depending on the number of slices, into smaller 2D data sets. However, the continuous modulation of the frequency introduces a small model error as projections aligned in  $k_z$ direction correspond to slighly different spectral positions. This error is negligible as long as the number of partitions is small compared to the total number of pulses applied.

After slice decoding, radial raw data from each slice is interpolated onto a 3-fold oversampled grid after gradient delay correction \cite{Block2011a}. The subspace-constrained reconstruction does not operate on raw data in the original or time domain $y_{t,k}$ but only requires raw data projected to the subspace $y_{p,k} =\mathcal{F}^H_t y_{t,k}$, which can be computed prior to gridding according to  $y_{p,k} =\sum_t W_{p,t} y_{t,k}$ with $W$ the low-frequency DFT matrix. In this preprocessing step, the summation can be performed prior to gridding (linearity of the Fourier transform) for raw data that has been sampled at different points $t$ in time at identical k-space positions $k$. This procedure lowers the computational burden for the gridding operation down to a single k-space per subspace coefficient and reduces the memory requirements tremendously.

Further reduction in computational complexity can be obtained by following the ideas of \cite{Tamir2016,Mani_2014}, namely taking advantage of the fact that pure temporal operators (such as the subspace transform) and pure spatial operates (such as the spatial Fourier transform or multiplication with coil sensitivity profiles) commute. The left-hand-side (LHS) appearing in the normal equation of the linear reconstruction problem with the given forward model (eq. 2) can then be formulated as

\begin{align}
\text{LHS} = S^{H}\underbrace{\mathcal{F}_{s}^{H}\mathcal{F}_{t}\mathcal{P}_{\vec{k}}\mathcal{F}_{t}^{-1}\mathcal{F}_{s}}_{=:{[\star]_{\text{TPSF}}}}Sx
\tag{3}
\end{align}

Radial sampling in the transform domain can be implemented efficiently by Fourier transforms and multiplication with a precomputed transform point spread function (TPSF) on a Cartesian grid similar to \cite{Uecker2010b}. This precomputation allows for solving the entire reconstruction problem in subspace without intermediate transformation to time domain.

The final reconstruction yields $P$ complex subspace coefficient maps. In principle, a SNR-optimal combination of these maps could be obtained by a coherent linear combination. This, however, requires exact knowledge about the individual phase relations in subspace for each pixel. Combining the images instead in a RSS manner, similar to coil combination in parallel imaging, is approximately optimal for high SNR pixels in parallel imaging \cite{Roemer_1990} and was used throughout this work as a post-processing step.

\subsection*{Water-Fat Separation}

In the context of bSSFP MRI, the difference in chemical shift of water and fat protons results in an additional frequency shift in their respective spectral responses. This shift is determined by the phase accrual per TR. Consequently, for an optimal water/fat separation ability, a TR has to be chosen such that water and fat spectra are maximally shifted with respect to each other, i.e. their respective bandings should be separated by $180^{\circ}$. Vasanawala and coworkers \cite{Vasanawala_2000} demonstrated that respective phase-cycled bSSFP data sets can be linearly recombined (linear-combined SSFP, LCSSFP) to yield virtual spectral responses designed to suppress water or fat, respectively. Here, we extend these ideas by exploiting the fact that the fmSSFP acquisition and reconstruction method yields full access to the spectral response. Similar to \cite{Vasanawala_2000}, a virtual spectral response is  approximated by a suitable linear combination of subspace coefficients (see Appendix). As weighting in Fourier subspace corresponds to a (cyclic) convolution in the spectral domain, this method extends the existing two-point method in \cite{Vasanawala_2000} in that signal intensities from all frequencies are taken into account to form the desired spectral response.

Additionally, in contrast to a (non-linear) RSS combination, the linear recombination of subspace coefficients supports partial volume signals in the same way as the linear subspace constraint does. This property is particularly important in the context of water-fat separation to avoid model mismatches.

\section*{Methods}

\subsection*{Simulation}

Numerical simulations have been implemented by the use of Extended Phase Graphs (EPG) \cite{Hennig2004}. For the special case of fully balanced sequences such as bSSFP and fmSSFP, transverse magnetization components do not experience any higher order dephasing from TR to TR, and, consequently, echo intensities can be calculated very efficiently using mere complex $3\times3$ matrix-vector multiplications. An efficient simulation framework was set up based on the freely available EPG-code written by Matthias Weigel \cite{Weigel2014a}.

\subsection*{MRI}

3D imaging was implemented by a fully-balanced stack-of-stars trajectory consisting of a set of aligned partitions. Individual partitions were radially sampled by repeating a sequence of uniformly distributed projections in linear order (turn-based pattern), while the overall temporal order of data acquisition was designed such that first a particular projection was acquired for each partition (partitions as innermost loop). This minimizes k-space "jumps", avoids eddy currents, and, most importantly, allows for smaller frequency modulation rates when compared to acquiring individual partitions sequentially. Frequency modulation according to eq. 1 was accomplished in such a way that the entire 3D acquisition covered one full frequency sweep ($\Delta \phi = 180^{\circ}$, $\Psi=360^{\circ}/N$). To drive the system into the desired dynamic steady state, preparatory scans prior to data acquisition were employed using the same phase modulation scheme as during data acquisition. This short preparation phase leads to an additional frequency offset in the dynamic steady state and has to be taken into account when referring to absolute frequencies. Note that only one single preparation phase is necessary as data acquisition in the dynamic steady-state does not require additional preparation phases in contrast to phase-cycled bSSFP.

In this work, the number of partitions was fixed to 40, and the number of different projection angles per partition was fixed to 101, which, for a fixed matrix size of $192\times192$, corresponds to a moderate undersampling factor. In the phantom and brain study, a voxel size of \SI{1x1x2}{\milli\meter} was realized with a repetition time of TR=\SI{5}{\milli\second} (TE=TR/2) and a total number of projections of $4\times101\times40$. In the knee study, a voxel size of \SI{1x1x3}{\milli\meter} was realized with a repetition time of TR=\SI{3.5}{\milli\second} (TE=TR/2) and a total number of projections of $8\times101\times40$. The employed flip angles were $\alpha=55^{\circ}$ (phantom study), and $\alpha=15^{\circ}$ (brain and knee study). Total acquisition times for 3D fmSSFP were 1 min 26 sec (phantom and brain), including a single initial preparation phase of 1000 pulses, and 2 min 4 sec (knee) including 3000 preparation pulses or dummy scans. Total acquisition times for 3D phase-cycled bSSFP were 1 min 41 sec (phantom and brain), including 4 individual preparation phases of 1000 pulses each. Note that for a turn-based acquisition pattern, every distinct k-space position was sampled as often as Fourier components were to be reconstructed (i.e. subspace size) to minimize aliasing.

Young adults without known illnesses were recruited and written informed consent was obtained prior to imaging according to the recommendations of the local ethics committee. MRI was performed at a field strength of 3T (Magnetom Prisma, Siemens Healthineers, Erlangen, Germany) with suitable elements of the standard 64-channel head coil for phantom and brain studies and with a 4-channel flex coil for the knee study.

\subsection*{Reconstruction}

Gridding, gradient delay correction, and precomputation of the TPSF was performed by custom MatLab routines, while image reconstruction was performed by a customized version of BART \cite{BART}   using ADMM \cite{Boyd_2010}  ($\rho=0.01$, 200 iterations) for optimization and a local low rank regularization constraint ($\lambda=0.001$ and block size $8\times8$). The multi-channel raw data was compressed to 10 virtual channels by principal component analysis \cite{Huang2008b} prior to image reconstruction. Coil sensitivity profiles were precomputed by an ESPIRIT reconstruction \cite{uecker2014espirit} using the density-compensated gridding solution of the 0th-order subspace coefficient as calibration data, and kept constant during optimization. Total reconstruction time for a single partition of a fmSSFP data set (subspace size $P=4$) was about \SI{80}{\s} on a 8-core Linux workstation, 64 GB RAM). For a subspace size $P$, reconstruction times of fmSSFP data sets are roughly increased by a factor of $P$ compared to an individual bSSFP data sets. Phase-cycled bSSFP images can be reconstructed independently while the fmSSFP subspace reconstruction requires a joint reconstruction of the coefficients. However, computational and memory demands of the proposed subspace-based approach depend only on subspace size and do not increase - in contrast to a naive implementation - when the total acquisition time of the fmSSFP experiment is increased.

Benkert and coworkers \cite{Benkert2015a} introduced with DYPR-SSFP a reconstruction technique for fmSSFP data that successfully removed banding artifacts by a conventional gridding reconstruction. This reference method was implemented here by density compensation, regridding of the radial raw data into a single frame, followed by spatial Fourier transform and a RSS combination across channels. In contrast to the original publication that employs a Golden-Angle-based trajectory, DYPR-SSFP in this work refers to the gridding solution regardless of the underlying trajectory.

\subsection*{Subspace}

As MRI measurements are typically contaminated by noise, the choice of the specific subspace and its size have also an effect on the noise amplification. While the Fourier transform as a unitary transform results in a constant noise level across subspace coefficients, an inherent trade-off between model error and noise amplification has to be made when choosing subspace sizes. In \cite{Tamir2016}, a comprehensive consideration of this effect can be found. Here, this analysis was adapted and applied to the low-frequency Fourier subspace in the presence of noise (compare eq. A10 in \cite{Tamir2016}).

\subsection*{Contrast}

The composition of RSS images introduces a new image contrast as signal intensities from the entire spectral response contribute in a non-linear way. An analytical derivation of this magnitude integration can be found in eq. A3 in the Appendix.

\subsection*{Water-Fat Separation}

For water/fat separation, a Hann-filtered sinc function approximating a box car with a spectral width of $180^{\circ}$ was used as a set of target coefficients ($C_p$ in eq. A2). A model fmSSFP signal was computed by EPG simulation assuming T1=\SI{0.4}{\second} and T2=\SI{0.1}{\second}. Approximate opposed-phase conditions for water and fat at the end of a repetition interval at a field strength of \SI{3}{T} were realized by choosing TR=\SI{3.5}{\ms}. For slab-selective excitation, sinc-shaped pulses with a duration of \SI{1.75}{\milli\second} and a bandwidth-time product of 6 were used. For the water/fat application a subspace size of $P=8$ and a sampling pattern consisting of 8 repeating turns was chosen.

For reference, a LCSSFP reconstruction \cite{Vasanawala_2000} was performed by linearily combining two phase-cycled bSSFP measurements ($\Delta\phi=0^\circ$ and the $\Delta\phi=180^\circ$) using the heuristic coefficients ${1,+i}$ and ${1,-i}$ to obtain a water and a fat image, respectively.

\section*{Results}

Figure 1 shows a comparison between the simulated fmSSFP signal response and the analytical spectrum of a bSSFP experiment. While the phases match closely, a characteristic asymmetry in the  fmSSFP signal can be seen in magnitude and complex representation. Figure 1d depicts the Fourier modes of the fmSSFP signal in comparison with the analytical expressions of the basic SSFP modes  (\cite{Ganter_2005,Zur1990}) as a function of the configuration order p. Despite the differences in the fmSSFP and bSSFP spectral responses, the relative error between fmSSFP and bSSFP does not exceed 4\% for the 8 lowest configuration modes. For illustration purposes, the simulated quadratic phase increment $\Psi=360/N=0.0891^{\circ}$ was chosen rather high compared to the measurements performed in this study where the total number of pulses $N$ was 4-fold (phantom and brain study) or 8-fold higher (knee study). The similarity in the spectral responses renders the lowest-order SSFP modes an excellent approximation of the Fourier modes of the fmSSFP signal.

Figure 2 shows the preserved energy fraction of a bSSFP signal as a function of the subspace size (upper row). The preserved energy fraction increases rapidly with increasing subspace size, and, for a fixed subspace size, compressability increases with longer T1, shorter T2, and larger flip angles. Constraining the subspace to size $P=1$ extracts only the DC part of the signal. This signal is not only obtained in a gradient-spoiled sequence where the signal distribution in each voxel is given by the bSSFP spectrum \cite{Hargreaves2012}, but also in fmSSFP MRI when all spectral data are pooled into a single k-space \cite{Benkert2015a,Benkert_2016}. With a subspace size of $P=8$, about 90\% of the signal energy is preserved for typical combinations of parameter values relevant in vivo which demonstrates that the low-frequency Fourier space is indeed a suitable subspace for bSSFP image reconstruction.

With increasing subspace size the model error decreases and the noise amplification increases. This bargain is visualized in the bottom row by the normalized root-mean-square error between the original, noiseless signal and its noisy subspace representation as a function of subspace size. Although optimal subspace sizes are parameter dependent, small mean-square errors are achievable with subspace sizes ranging up to $P=8$.

Possible aliasing effects can be appraised by analyzing the dependence of the TPSF on the sampling trajectory. To this end, the TPSF of a turn-based (Fig. 3) and Golden-Angle based sampling scheme (Fig. 4) is shown. The undersampling in k-t-space results in aliasing between subspace coefficients (off-diagonal contributions) and in missing projections for individual subspace coefficients. The chosen sampling pattern determines the balance between these two effects. When sampling with a continuous Golden-Angle scheme, overlap in k-space is minimized and each subspace coefficient exhibits a close-to-uniform coverage in k-space. In this case, however, incoherent temporal aliasing affects all subspace coefficients. Sampling according to a turn-based pattern leaves a certain fraction of projection angles uncovered and therefore increases the undersampling in each subspace coefficient's k-space. On the other hand, due to the repetitive nature of the turn-based pattern, each sampled projection angle is covered several times and through-time aliasing is reduced such that low temporal frequencies in the sampled signal appear free of aliasing up to an order that is given by the number of full turns sampled. 

Throughout this work, the turn-based pattern was utilized to reconstruct images from fmSSFP measurements, demonstrated in Fig. 5 for a doped water phantom where one of the shimming coils was used to create an approximately linear off-resonance gradient across the FoV. The four phase-cycled bSSFP images are depicted along with their RSS reconstruction (5a). The four lowest modes of the fmSSFP data set are reconstructed and shown in magnitude and phase representation along with their RSS combination (5b). For comparison, matching synthetic images from the fmSSFP reconstruction are generated. The RSS reconstructions clearly demonstrate the effective removal of banding artifacts by the proposed subspace-based fmSSFP acquisition and reconstruction technique. Note that the fmSSFP data set only required one initial preparation phase (here \SI{5}{\s})  while the four bSSFP data sets required four individual preparations (\SI{20}{\s} in total).

Figure 6 illustrates the parameter dependencies (contrast) of the RSS reconstruction. For comparison, the signal intensity of the on-resonant bSSFP signal ($\Delta \phi = 180^{\circ}$) is plotted as well. The functional dependencies of the two signals are generally very similar, so that the integrated spectral energy exhibits the same characteristic T2/T1 contrast of classic bSSFP images. However, the flip angle dependency reveals that the optimal flip angle for the integrated contrast is smaller than its bSSFP equivalent of $\alpha_{\text{opt}}\approx\arccos\frac{T1/T2-1}{T1/T2+1}$ \cite{Sekihara_1987} and that in the regime of very small flip angles still a substantial signal intensity can be achieved due to intense off-resonant contributions.

Figure 7 shows exemplary transversal partitions of a 3D MRI measurement of a human brain. Composite RSS images have been computed from four phase-cycled data sets and from the four lowest-frequency modes for fmSSFP in analogy to the phantom study. The overall appearance of brain tissue and cerebrospinal fluid is very similar. In close analogy to the phantom study, residual banding artifacts remain in phase-cycled bSSFP images (arrows) in contrast to fmSSFP images in which no residual bandings can be found. Under conditions of good magnetic field homogeneity (a), bSSFP and fmSSFP images are free of bandings and exhibit very similar contrast independent of the chosen sampling scheme. However, in the presence of a strong linear field gradient, a Golden-Angle-based sampling scheme leads to intense noise-like artifacts in the subspace-constrained fmSSFP reconstruction. This finding is in agreement with the corresponding TPSF that allows incoherent aliasing across all subspace coefficients.

Figure 8 depicts reconstructions from a 3D MRI measurement that has been performed at the human left knee joint with sagittal slice encoding. Besides anatomical images, water-fat separated images are obtained by linear recombination of the subspace coefficients of the fmSSFP reconstruction according to eq. A1. The presented partition shows the patellofemoral and tibiofemoral joint. Articular cartilage and synovial fluids appear bright in the water image (LCfmSSFP, arrows) and hypointense in the fat image. Bone marrow and subcutaneous fat appear with high signal in the fat image. Reconstructed spectra (8c) of the fmSSFP measurement (dashed lines) and after recombination (solid lines) are shown exemplarily for a voxel in the bone marrow (red) and for a voxel in the synovial fluid (blue). The approximate opposed-phase condition can be seen in the shift of the two minima in the fmSSFP spectrum. After recombination, two similarly shaped virtual spectra can be obtained, where water and fat images correspond to the off-resonant positions of $270^\circ+17^\circ=287^\circ$ and $90^\circ+17^\circ=107^\circ$, respectively, as the additional frequency offset of $17^\circ$ due to the short preparation phase has to be taken into account. The reference method LCSSFP suffers from reduced SNR in water and fat images as only two phase-cycled bSSFP data sets were combined. Additionally, streak artifacts with increased intensity as well as water/fat swaps (LCSSFP, arrows) can be seen that are absent in the proposed LCfmSSFP method.

\newcommand{\figscale}{0.35}
\begin{figure}[htbp]
\centering
\includegraphics[width=\textwidth]{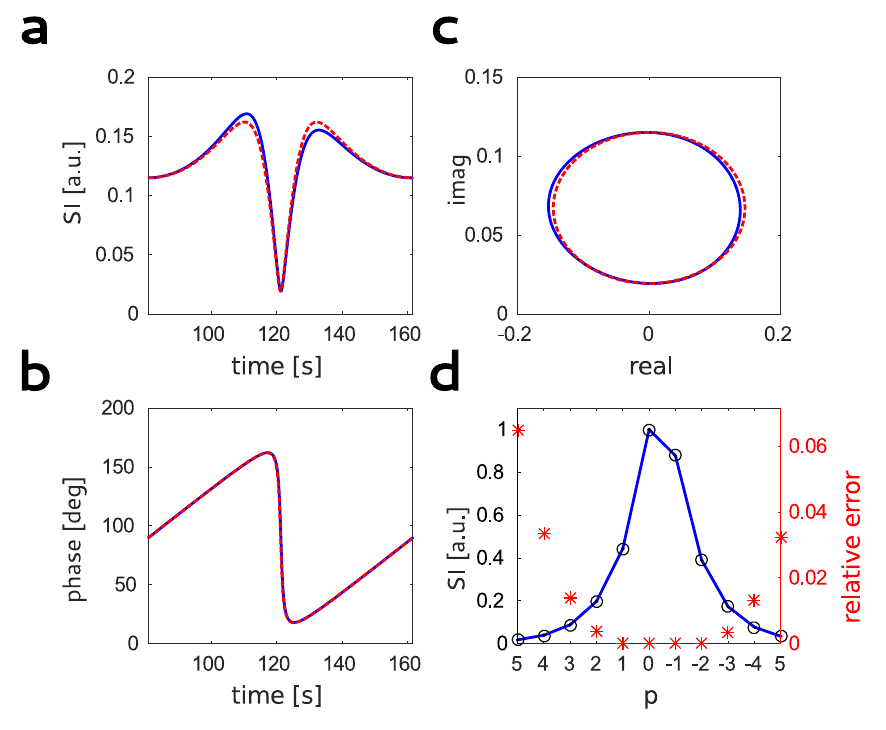}
\caption{\label{Figure 1}
Comparison between the EPG-simulated fmSSFP signal response (solid blue) and the analytical spectrum of a bSSFP experiment (dashed red) represented as magnitude and phase time courses (a-b) and geometrically in the complex plane (c). Fourier modes of the fmSSFP signal (d, solid black) are shown in comparison with the analytical expressions of the basic SSFP modes (open circle) \cite{Ganter_2005,Zur1990} as a function of configuration order p. Additionally, the relative error between the Fourier modes of the fmSSFP and the analytical expressions of the basic SSFP Fourier modes is depicted in red. The x-axis has been inverted to be consistent with the order notation as introduced in \cite{Heule2014}. The numerical simulation was performed for a total number of $4\times101\times40$ pulses after an initial preparation with T1=\SI{1}{\s}, T2=\SI{0.1}{\s}, TR=\SI{5}{\ms}, and $\alpha=15^{\circ}$.
}
\end{figure}

\begin{figure}[htbp]
\centering
\includegraphics[width=\textwidth]{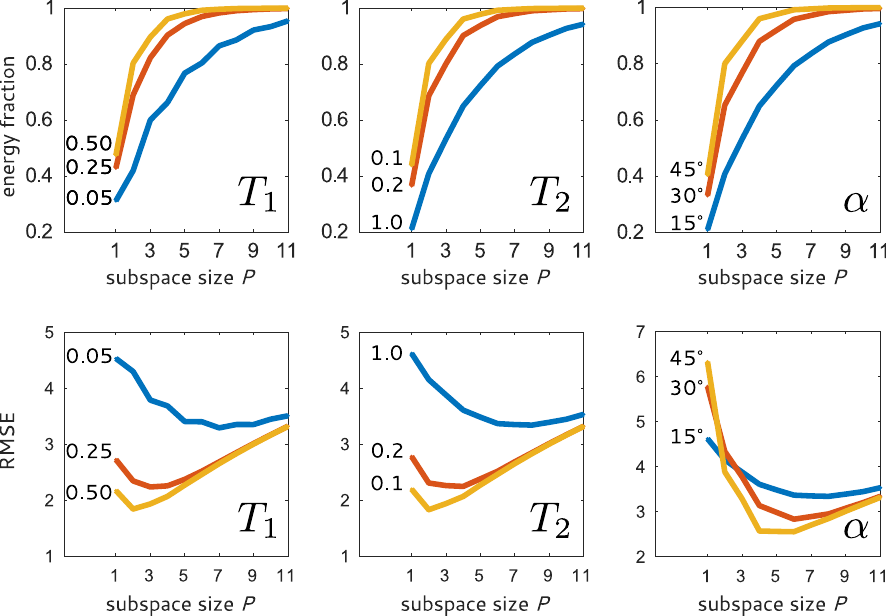}
\caption{\label{Figure 2}
Top row: The preserved energy fraction of a bSSFP signal as a function of subspace size for three different values of T1, T2, and flip angle, while keeping the respective other parameters constant. Energy fractions were computed by dividing the energy of the p-th order subspace approximation by the energy of the full spectral response. The preserved energy fraction increases rapidly with increasing subspace size for all parameter combination. For a fixed subspace size, compressibility increases with longer T1, shorter T2, and larger flip angles.
Bottom row: Normalized root-mean-square error (compare eq. A10 in \cite{Tamir2016}) as a function of subspace size for the same values of T1, T2, and flip angle. For the calculations, a constant noise level of $\sigma=5\%$ of the maximal possible signal level was assumed for each time point.
Left column: T2=\SI{50}{\ms}, $\alpha=15^{\circ}$, center column: T1=\SI{1}{\s}, $\alpha=15^{\circ}$, right column: T1=T2=\SI{1}{\s} 
}
\end{figure}

\begin{figure}[htbp]
\centering
\includegraphics[width=\textwidth]{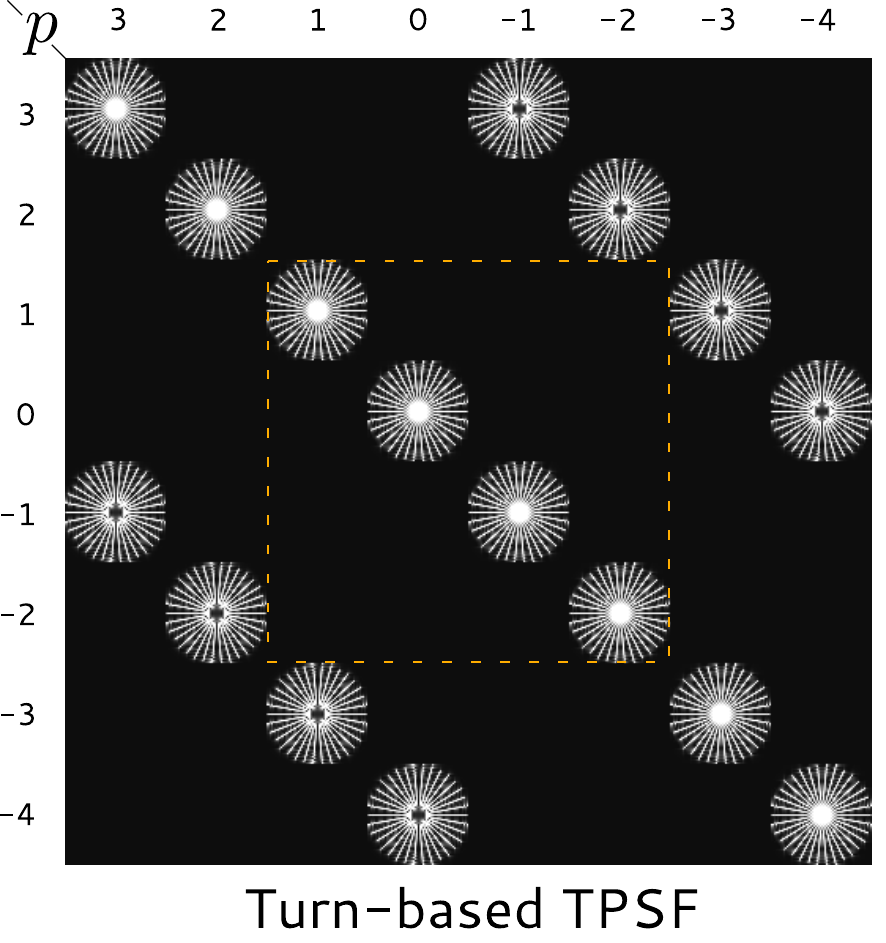}
\caption{\label{Figure 3}
Fourier transform of the TPSF for a turn-based undersampling scheme with 4 repetitions of 17 uniform projections. right: Golden-Angle pattern with $68=4\times17$ total projections). The dotted square (left) indicates a subspace size of $P=4$ for which no aliasing between the reconstructed coefficients occurs. The maximum size of this subspace is given by the number of repetitions for the turn-based pattern.
}
\end{figure}

\begin{figure}[htbp]
\centering
\includegraphics[width=\textwidth]{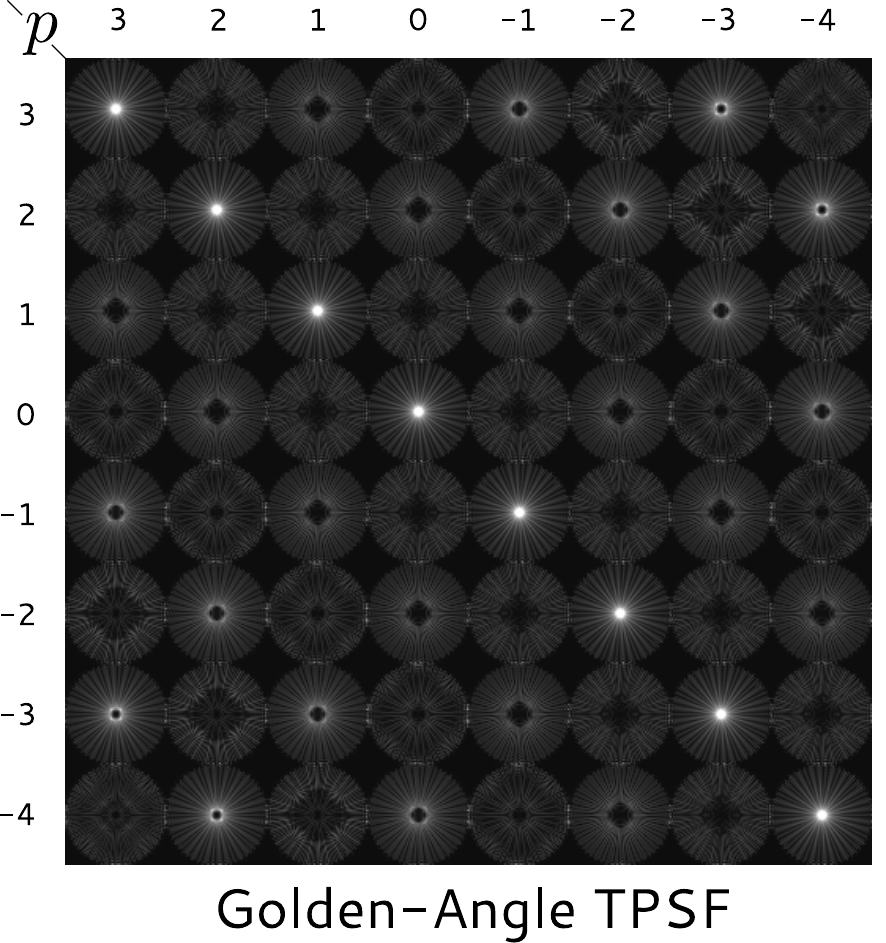}
\caption{\label{Figure 4}
Fourier transform of the TPSF for a Golden-Angle undersampling scheme with $68=4\times17$ total projections. Undersampling in k-t-space results in aliasing between subspace coefficients (off-diagonal contributions) and in missing projections for individual subspace coefficients. While the turn-based pattern results in a combination of absent and strongly coherent aliasing (see Fig. 3), the Golden-Angle scheme produces incoherent aliasing across all subspace coefficients.
}
\end{figure}

\begin{figure}[htbp]
\centering
\includegraphics[width=\textwidth]{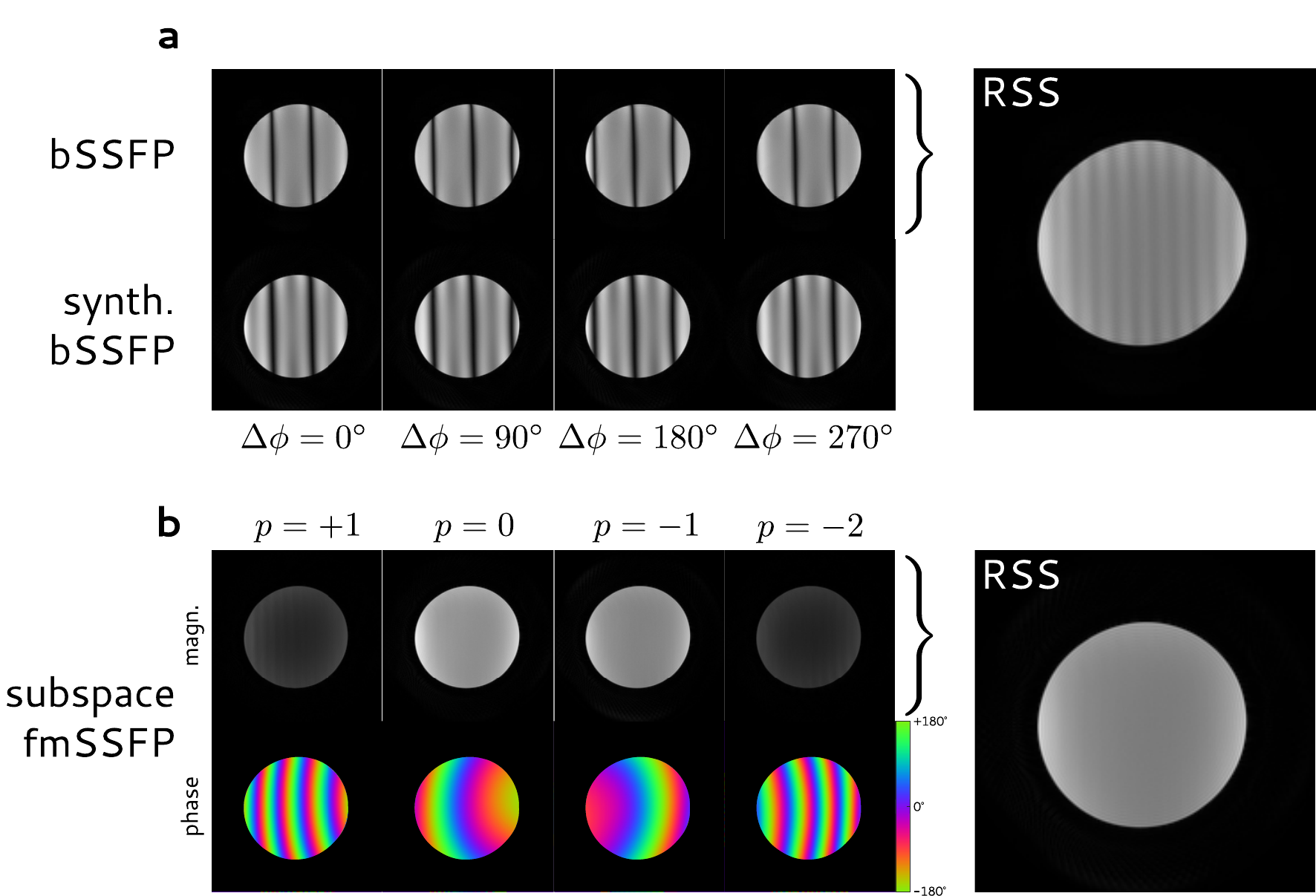}
\caption{\label{Figure 5}
Comparison between image reconstructions from four phase-cycled bSSFP data sets and a corresponding fmSSFP data set. Depicted are the four individual phase-cycled bSSFP images and the corresponding RSS reconstruction (a). For the fmSSFP data set, the four lowest order modes are reconstructed and shown in magnitude and phase representation  along with their RSS combination (b). For improved visualization, the phase maps have been masked to the support of the object. For comparison, synthetic images from the fmSSFP reconstruction are generated that match the frequency offsets of the phase-cycled measurements. The RSS reconstruction of the bSSFP images still exhibits residual banding artifacts of substantial intensity while in the corresponding fmSSFP reconstruction these artifacts appear hardly perceivable. For an unbiased comparison, precomputed coil sensitivity profiles for bSSFP reconstruction were taken from fmSSFP. 
}
\end{figure}

\begin{figure}[htbp]
\centering
\includegraphics[width=\textwidth]{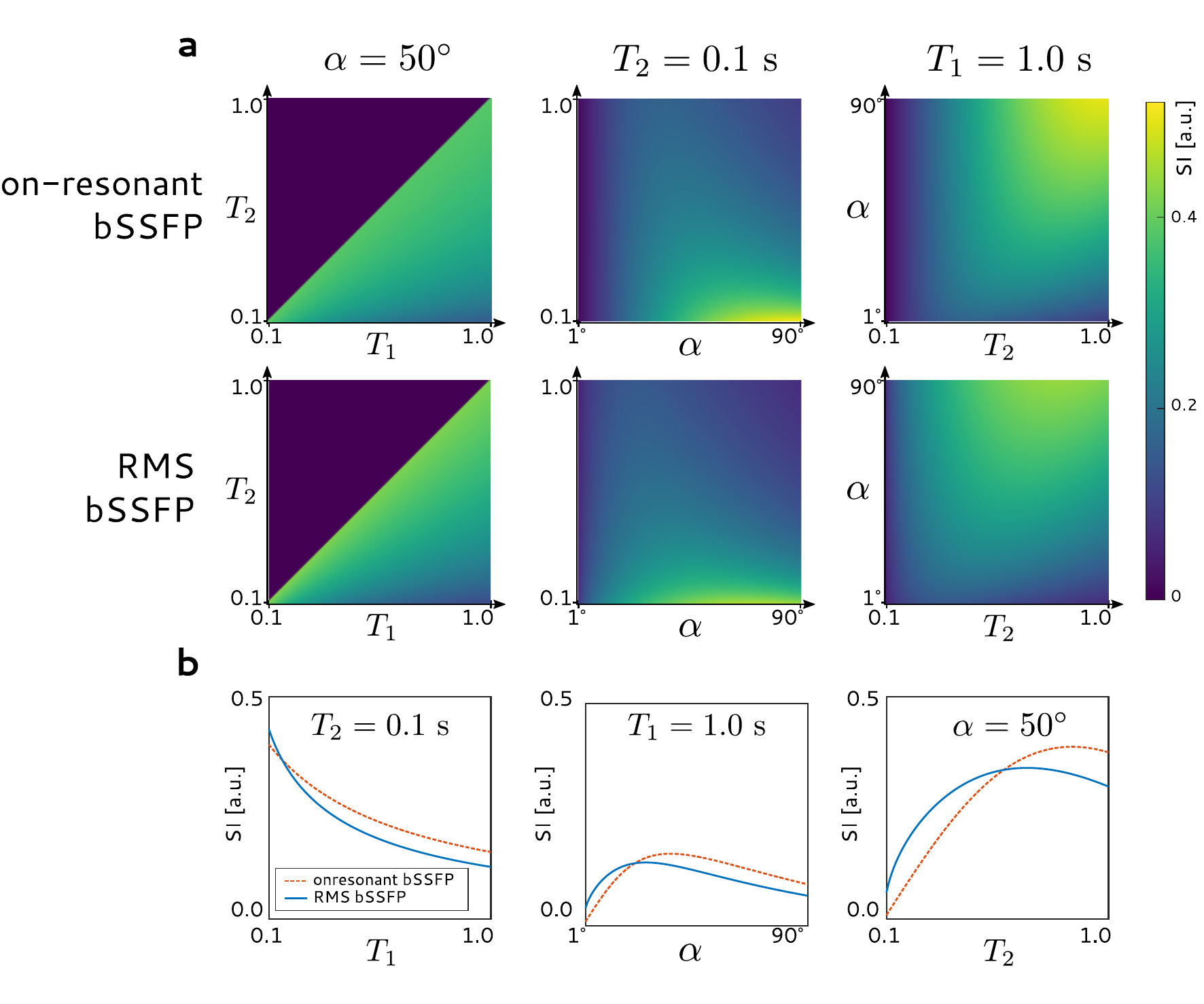}
\caption{\label{Figure 6}
Parameter dependency of the on-resonant bSSFP signal ($\Delta \phi = 180^{\circ}$) in comparison to the root-mean-square (RMS) of the full signal's energy (eq. A3). The parameters alpha, T2, and T1 were individually kept constant  while varying the respective other two parameters (a). For improved visualization, line plots for selected parameter configurations are additionally displayed (b, corresponding to horizontal profiles in the upper graphs). General similarity in parameter dependency explains the same characteristic T2/T1 contrast observed for subspace fmSSFP and on-resonant bSSFP.
}
\end{figure}

\begin{figure}[htbp]
\centering
\includegraphics[width=0.9\textwidth]{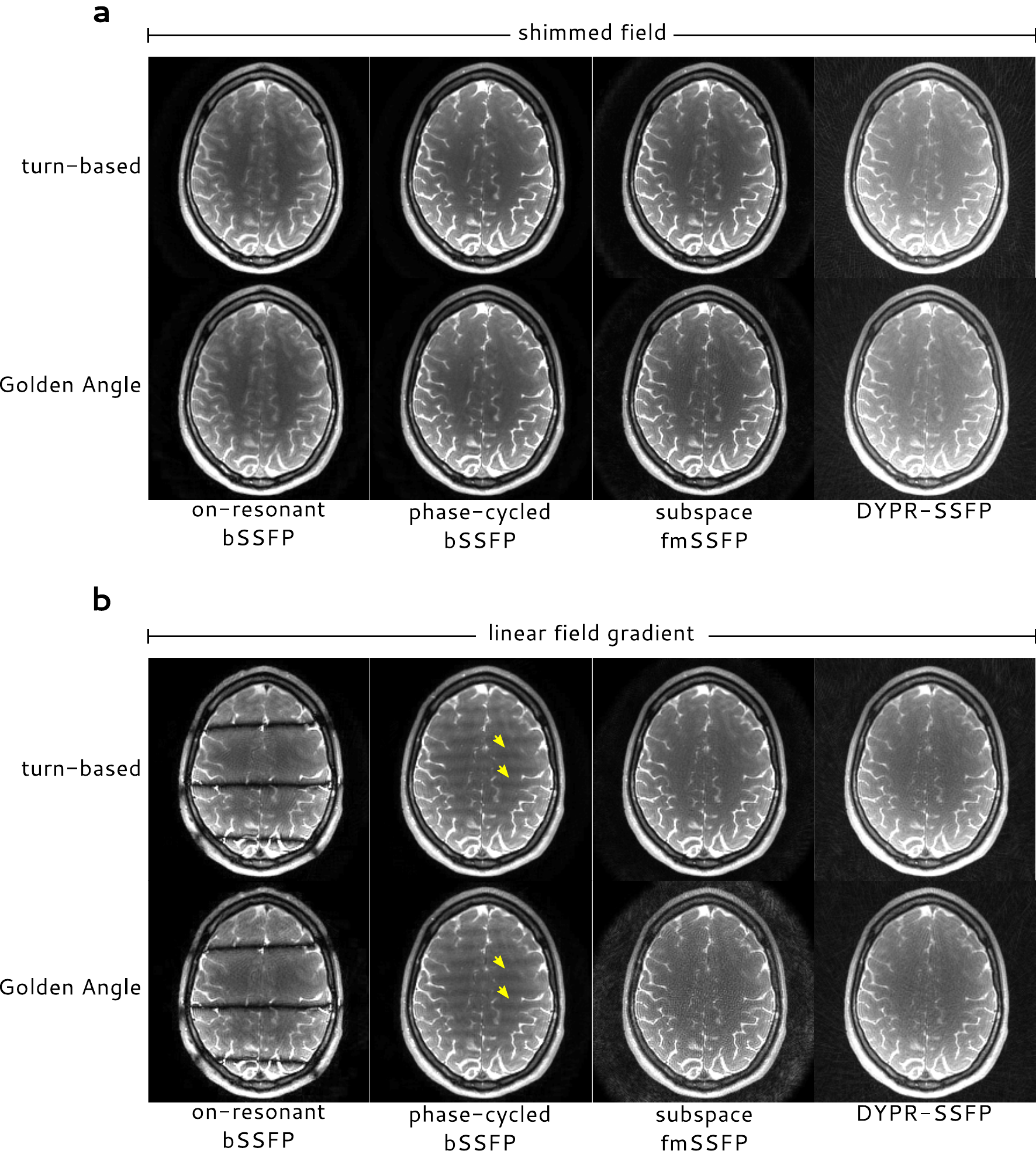}
\caption{\label{Figure 7}
Transversal partitions of a 3D MRI measurement of a human brain. In (a), imaging in a well shimmed static field was performed using turn-based and Golden-Angle-based sampling schemes. Note the loss of contrast between gray and white matter for the DYPR-SSFP reconstruction. In (b), the same type of acquisition and reconstructions are shown in the presence of a strong linear field gradient across the FoV. The overall appearance of brain tissue and fluid in phase-cycled bSSFP and fmSSFP images is very similar. In close analogy to the phantom study, residual banding artifacts remain in phase-cycled bSSFP images (arrows) in all brain regions in contrast to fmSSFP images where no residual bandings can be found. Only minor differences can be seen between the two employed trajectories except for the subspace-based fmSSFP reconstruction in which the Golden-Angle reconstruction exhibits strongly increased noise-like artifacts. For an unbiased comparison, precomputed coil sensitivity profiles were taken from fmSSFP for all reconstructions.
}
\end{figure}

\begin{figure}[htbp]
\centering
\includegraphics[width=\textwidth]{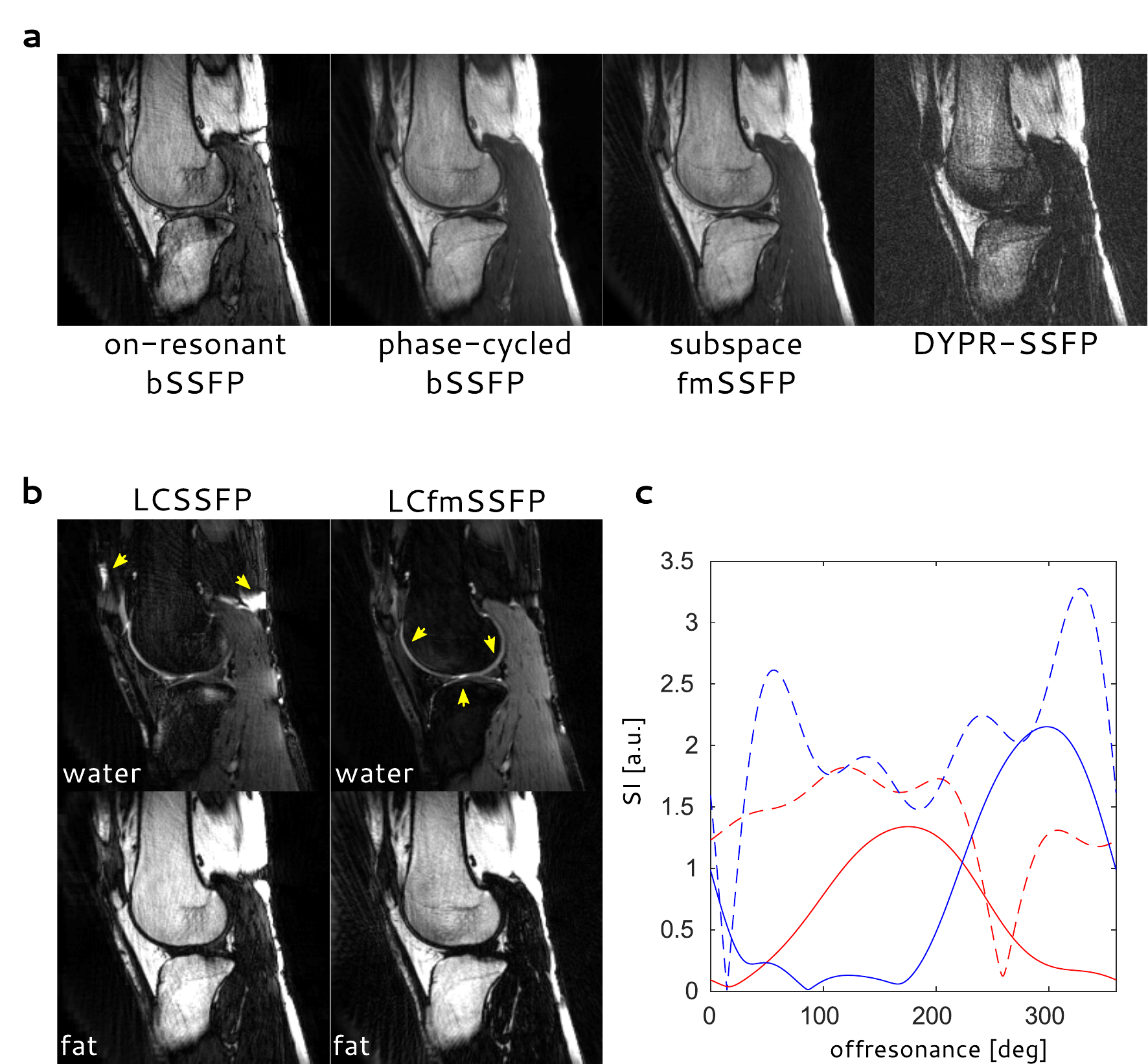}
\caption{\label{Figure 8}
Sagittal partitions of a 3D MRI measurement of a left human knee. Shown are the anatomical images (a) along with water/fat separated images (b). The presented partition shows the patellofemoral and the tibiofemoral joint. Articular cartilage and synovial fluids appear bright in the water image (LCfmSSFP, arrows) and hypointense in the fat image. Reconstructed spectra (c) of the fmSSFP measurement (dashed lines) and after linear recombination (solid lines) are shown exemplarily for a voxel in the bone marrow (red) and for voxel in the synovial fluid (blue). After recombination, two similarly shaped virtual spectra are obtained, where water and fat images correspond to off-resonant positions of $287^\circ$ and $107^\circ$, respectively.
The LCSSFP water and fat images suffer from reduced SNR as only two phase-cycled bSSFP data sets were combined. Additionally, streak artifacts with increased intensity as well as water/fat swaps (LCSSFP, arrows) can be seen that are absent in the proposed LCfmSSFP method.
}
\end{figure}

\section*{Discussion}

Recently, several methods have been proposed to accelerate phase-cycled bSSFP measurements \cite{Hilbert_2017,Cukur_2015,Biyik_2017,Ilicak_2016,Shcherbakova_2017,_ukur_2009,Bj_rk_2013}. Individual measurements exhibit a large degree of redundancy which can be exploited by compressed sensing techniques  \cite{Lustig_2007,Donoho_2006} to enable severe undersampling. Hilbert and coworkers \cite{Hilbert_2017} proposed a model-based reconstruction that reduces the number of unknowns in the reconstruction problem by estimating the relaxation time ratio $\Gamma={T_2}/{T_1}$  instead of signal intensities. However, independent of the degree of undersampling, a certain portion of the total acquisition time is dedicated to preparation phases to ensure that the individual steady states have been reached. The specific amount of time $t_\text{prep}$ spent in such preparation phases varies in the literature from $t_{\text{prep}}$=\SI{4.9}{\s} at \SI{356}{s} total acquisition time ($t_{\text{ACQ}}$) \cite{Hilbert_2017}, to $t_{\text{prep}}/t_{\text{ACQ}}=\SI{16}{\s}/\SI{1139}{\s} $ \cite{Benkert2015a} , and up to $t_{\text{prep}}/t_{\text{ACQ}}=\SI{163}{\s}/\SI{196}{\s}$   \cite{Bangerter_2004}, with shorter preparation phases jeopardizing steady-state conditions and longer preparation phases limiting the effectivity of bSSFP.

Both methods, fmSSFP and bSSFP, require an initial preparation phase, which could potentially be shortened. Apart from the common $\alpha$/2-TR/2 preparation \cite{deimling1994} a variety of schemes is possible \cite{bieri2013} to catalyze the initial preparation phase faster than simple dummy scans. In this work, the focus lies on the removal of intermediate preparation phases and consequently no attempts have been made to implement more sophisticated preparation schemes for the initial phase so far.

This work presents a subspace-based method for the reconstruction of banding-free images from fmSSFP data sets. The reconstruction of 3D fmSSFP MRI data has been demonstrated for a phantom as well as for the human brain and knee. In all cases, the reconstructed images appear free of banding artifacts, show similar image contrast when compared to phase-cycled bSSFP data, and were acquired in shorter time by waiving intermediate preparation phases. Here, the combination of fmSSFP with 3D MRI and its high number of excitations allows for the use of small modulation rates that yield signal responses close to classic bSSFP spectra. In addition, approximation errors are shown to decrease with decreasing configuration order, and by choosing the low-frequency Fourier transform as a subspace, errors introduced by approximating fmSSFP signals with analytical expressions for bSSFP are further reduced. Despite this close match, the non-linear combination of subspace coefficients introduces a novel image contrast potentially deviating from classic bSSFP. To investigate the differences, an analytical expression for the magnitude integration of the complex bSSFP spectrum was derived that demonstrates the same characteristic T2/T1 dependency (contrast) as classic bSSFP.

For fmSSFP, the minimum number of measurements required for a banding-free image depends on the rate of signal decay in Fourier space. With slowly decaying signals and small subspaces, not only energy is lost in the reconstruction, but also aliasing from higher order Fourier components (above Nyquist frequency) can deter image quality. The situation for phase-cycled bSSFP is different in that temporal aliasing cannot occur as a non-dynamic steady-state is sampled. However, residual banding artifacts are more intense as the spectral response is sampled at few frequencies only. As a consequence, there is no one to one correspondence between the number of phase cycles for artifact-free reconstruction in fmSSFP and bSSFP. Residual aliasing, bandings and generated contrast are complex functions of flip angle, relaxation times, and sampling trajectory. 

In \cite{Tamir2016}, Tamir and coworkers present a subspace-based acquisition and reconstruction method for 3D FSE imaging. In contrast to the conventional center-out view ordering, a randomized (shuffled) view ordering is employed which spreads interferences between subspace coefficients incoherently and allows for the reconstruction of sharper images. A similar effect of the (under)sampling trajectory is also observed in this work. Here, the Fourier subspace properties also result in incoherent aliasing between individual subspace coefficients when choosing a quasi-random view-ordering such as a Golden-Angle scheme. By the repetition of a single-turn pattern these aliasing artifacts can be reduced at the expense of undersampled k-spaces for all individual subspace coefficients. However, parallel imaging methods can compensate for this particular undersampling by exploiting additional spatial information encoded in the coil sensitivity profiles \cite{Pruessmann1999,Griswold2002,Uecker_2013}, while temporal aliasing across different subspace coefficients cannot be resolved by additional spatial information. Turn-based sampling is therefore favored over Golden-Angle sampling when combining the proposed subspace method with parallel imaging. Consequently, we propose as a general procedure to first choose a suitable subspace size  $P$, and second, based on that choice, to repeat a single-turn pattern $P$ times.

Benkert and coworkers recently presented a fast imaging technique named DYPR-SSFP \cite{Benkert2015a,Benkert_2016} that is also based on radial sampling and fmSSFP. Similar to our approach, DYPR-SSFP also requires only a single preparation phase and performs data sampling in the dynamic steady state. However, DYPR-SSFP does not try to reconstruct the spectral information of the fmSSFP signal. Instead, only one single k-space is reconstructed which comprises all projections that are acquired in a Golden-Angle view order. Although a complete removal of banding artifacts could also be achieved in this way, no attempts have been made to reconstruct the spectral content of the fmSSFP signal as only the DC signal component is captured when pooling data from the entire phase cycle into one single k-space. Our proposed subspace-approach can thus be considered an extension in which the subspace size $P=1$ corresponds to the special case of the DYPR-SSFP technique and larger subspaces allow for a reconstruction of the full spectral content of the generated fmSSFP signal.

Additionally, the DC signal component reconstructed in DYPR-SSFP comes at reduced signal strength when compared to the standard on-resonant bSSFP profile (see Fig. 2 and Fig. 5) because the complex average of the spectral response ignores phase relations and leads to signal cancellation due to dephasing effects. To overcome these effects and to improve SNR, Slawig and coworkers proposed a multi-frequency reconstruction technique for fmSSFP \cite{Slawig_2017} that comprises phase correction, multiple reconstructions for different assumed frequencies, and a final maximum intensity projection. Instead of relying on separate reconstructions for different off-resonance frequencies, our method reconstructs a single low-frequency response where the off-resonance present during acquisition translates into a linear phase term in the subspace.

In this work, the bSSFP-based water/fat separation method LCSSFP \cite{Vasanawala_2000} has been extended in such a way that the entire spectral information is linearly combined to form virtual responses. We demonstrated the water-fat separation ability in the human knee with a subspace size of $P=8$. As the utilized set of coefficients are computed based on a model signal, the separation ability, just like LCSSFP, depends on the deviation between tissue relaxation parameters and assumed parameters. These dependencies have not been analyzed and demand further investigations. Additionally, fmSSFP-based water/fat separation is sensitive to off-resonance effects as caused by imperfect shimming. These off-resonances can lead to incorrect assignments of water and fat contributions (water/fat swaps). Typically, these shortcomings can be overcome by a proper estimation of the background field map which is beyond the scope of this work.

Stack-of-stars sampling schemes using complementary samples recently have been shown to outperform schemes with the same samples in each partition \cite{Rosenzweig_2017,zhou2017golden}. In this work, the same samples in each partition were used to enable decoupling of the 3D problem by simple inverse Fourier transform along slice dimension. The main limitation of using complementary samples in the proposed approach is the increase in memory demand as the full 3D problem has to be solved as a whole in subspace including prior estimation of 3D coil sensitivity profiles. This and a possible joint estimation of coil sensitivities and subspace coefficients similar to \cite{Uecker2008} will be part of future work.

\section*{Conclusions}

With the novel subspace-based fmSSFP MRI technique, a time-efficient 
alternative to phase-cycled bSSFP is presented. Banding-free images can be 
reconstructed without the need for excessive preparation phases. The subspace 
formulation of the imaging problem allows for a computationally efficient 
reconstruction of 3D data sets and a RSS combination yields high-SNR images in 
a simple post-processing step preserving the characteristic bSSFP contrast. 
Access to the spectral information of the fmSSFP signal was exploited and 
water/fat separation exemplarily demonstrated as a possible application of the 
presented approach.

\section*{Acknowledgement}

The authors would like to thank Dr. Dirk Voit for his continuous support in sequence development and Dr. Klaus-Dietmar Merboldt for his valuable advice in knee joint MRI.

\section*{Appendix}

\subsection*{Linear-combination fmSSFP (LCfmSSFP)}

Like in LCSSFP, the aim of LCfmSSFP is to generate a certain target response $f(\theta)$ by linearly recombining measured signals using a set of complex coefficients $w_{p}$:

\begin{align}
f(\theta)=\sum_{p=-P/2}^{+P/2-1}w_{p}x_{p}e^{ip\theta}
\tag{A1}
\end{align}

where $s(\theta)=\sum_{p=-P/2}^{+P/2-1}x_{p}e^{ip\theta}$  is the given subspace representation of a signal.

These coefficients can be found by approximating the desired spectral response by a low-frequency Fourier expansion of the form

\begin{align}
f(\theta)\approx\sum_{p=-P/2}^{+P/2-1}C_{p}e^{ip\theta}
\tag{A2}
\end{align}

The desired coefficients can then be obtained by simply setting $w_{p}:=C_{p}/x_{p}$ as $\sum_{p=-P/2}^{+P/2-1}w_{p}x_{p}e^{ip\theta}=\sum_{p=-P/2}^{+P/2-1}C_{p}e^{ip\theta}\approx f(\theta)$ holds.

With these coefficients, any subspace representation of an actual fmSSFP spectral profile can be transform  into a virtual spectral response. Note that these coefficients $w_{p}$ are functions of T1, T2, and flip angle of the model signal. Like in LCSSFP, other parameter combinations might give spectral responses that deviate from the target response, but linearity of the subspace transformation ensures proportionality to water and fat content, respectively.

\subsection*{Contrast}

The complex bSSFP signal right after the RF pulse can be described as (see e.g. \cite{Shcherbakova_2017}):

$$S(\theta)=M\frac{1-ae^{i\theta}}{1-b\cos\theta}\ ,$$\\
where $M = \frac{M_{0}(1-E_{1})\sin\alpha}{1-E_{1}\cos\alpha-E_{2}^{2}(E_{1}-\cos\alpha)} $, $a  =  E_{2} $,\\
$b  =  \frac{E_{2}(1-E_{1})(1+\cos\alpha)}{1-E_{1}\cos\alpha-E_{2}^{2}(E_{1}-\cos\alpha)} $, $E_{1}  =  e^{-\frac{\text{TR}}{T_{1}}} $, $E_{2}  =  e^{-\frac{\text{TR}}{T_{2}}}$

\vspace*{2\baselineskip}
As the energy in the Fourier domain equals the energy in the spectral domain, the RSS operation on the subspace coefficients approximates the root-mean-square of the energy $E$ of the complex bSSFP spectrum:

$$E=\frac{1}{2\pi}\int_{-\pi}^{+\pi}\text{d}\phi|S(\theta)|^2=\frac{1}{2\pi }|M|^2\left[(1+a^2)\beta -2a\gamma \right]$$
where $\beta=\int_{-\pi}^{+\pi}\text{d}\phi\frac{1}{(1-b\cos\theta)^{2}}$ and $\gamma=\int_{-\pi}^{+\pi}\text{d}\phi\frac{\cos\theta}{(1-b\cos\theta)^{2}}$

With the analytical solutions of these two integrals [see \cite{ZwillingenDaniel20140918}, pp. 404, 407]
$\beta = \frac{2\pi}{\left(1-b^{2}\right)^{\frac{3}{2}}}$ and $\gamma = \frac{2b\pi}{\left(1-b^{2}\right)^{\frac{3}{2}}}$
the final root-mean-square is given by

\begin{align}
\sqrt{E}=|M|\sqrt{\left(1-b^2\right)^{-\frac{3}{2}}\left[1+a^2-2ab\right]}
\tag{A3}
\end{align}

\bibliographystyle{mrm}
\bibliography{bibliography}

\iftoggle{MRM}{
\newpage
\begin{NoHyper}
\paragraph{\Cref{Figure 1}:}	\nameref{Figure 1}.
\paragraph{\Cref{Figure 2}:}	\nameref{Figure 2}.
\paragraph{\Cref{Figure 3}:}	\nameref{Figure 3}.
\paragraph{\Cref{Figure 4}:}	\nameref{Figure 4}.
\paragraph{\Cref{Figure 5}:}	\nameref{Figure 5}.
\paragraph{\Cref{Figure 6}:}	\nameref{Figure 6}.
\paragraph{\Cref{Figure 7}:}	\nameref{Figure 7}.
\paragraph{\Cref{Figure 8}:}	\nameref{Figure 8}.
\end{NoHyper}
}

\end{document}